The PCL Framework: A strategic approach to comprehensive risk management

in response to climate change impacts

Youssef Nassef[1]

## 1. Introduction

In recent years, climate-related extreme events have resulted in losses that have consistently exceeded prior estimates, and surpassed the capacity for planned response that had been set up for such eventualities. This situation has increasingly strained government capacities in serving as "insurers of last resort". This has held true in both developed and developing countries, and has revealed a gap in the efficacy of existing climate risk management approaches, and associated economic tools, to avert loss of life and human suffering, and to minimize destruction and loss of material assets.

Numerous approaches, methods and tools have been proposed to guide the process of risk management in response to climate risks. In many cases, these approaches have focused on one or more of the following:

- risk layering to address financial risk, typically mapped to severity or loss magnitude, or to probability, frequency or return periods;[2]
- disasters or extreme events, rather than the full range of climate change impacts; and
- maximizing risk reduction or preemptive adaptation, while assigning residual or excess risk to ex-post remedial action, including through altruistic interventions.

Such approaches contend that low-severity/high-frequency events be addressed through risk reduction measures, while higher severity and lower frequency events be addressed through contingent finance and insurance. The highest category of severity is deemed to be best addressed through humanitarian response and relief operations, or other ad-hoc support systems.

The underlying reasoning of such methodologies has been the need to first prioritize all preemptive actions that reduce risks as far as economically possible,[3] and then find contingent measures to deal with what is deemed to be "residual risks". Whatever is left over from these becomes the purview of humanitarian intervention. There is, therefore, an inherent hierarchy where preemptive action is seen to be the most desirable type of response, followed by contingent action, followed by humanitarian or similar interventions to address whatever losses are left over from the climate impacts under consideration.

---

[1] Youssef Nassef serves as Director of the Adaptation Programme at the Secretariat of the United Nations Framework Convention on Climate Change UNFCCC). Views presented in this article are of the author and do not necessarily reflect the position of the United Nations or the UNFCCC.
[2] See, for example, Mechler R, Bouwer LM, Linnerooth‑Bayer J, Hochrainer‑Stigler S, Aerts JCJH, Surminski S, Williges K (2014). Managing unnatural disaster risk from climate extremes. Nature Climate Change 4(4):235–237.
[3] UNFCCC. Technical Paper: Mechanisms to manage financial risks from direct impacts of climate change in developing countries. FCCC/TP/2008/9. UNFCCC: 2008.

One problem with this implicit hierarchy is its inconsistency with the need to evolve towards a comprehensive approach that takes account, and optimizes the use, of all possible responses, including loss acceptance where most appropriate. The approach would also need to take account of societal values in assessing loss tolerability.

In addition, current approaches have failed to make the case for sufficient investment in preemptive and contingent measures in response to climate risk. At the same time, humanitarian response and relief efforts are also often insufficient in effectively and fully addressing, in a timely manner, the residual losses arising from climate impacts.[4] Recent advances in risk quantification have spurred the potential for developing a rigorous step-wise method for optimized comprehensive risk management, including through the use of quantitative predictive models and use of Geographic Information Systems and remotely sensed data, complemented by the evolving prospects of big data analytics, the resurgence of artificial intelligence and the Internet of Things.[5] These advances present opportunities for reducing information asymmetries and for employing predictive analytics in new ways that can inject efficiencies in the world's adaptive response to climate change.[6]

This work presents a systematic approach to address these new realities, resolve the abovementioned concerns, and transcend existing efforts in a way that is both climate-change and policy relevant. It makes a case for the optimization of response action across the three main response clusters, namely preemptive adaptation (P) or risk reduction; contingent arrangements (C); and loss acceptance (L), without a predetermined hierarchy across them. The "PCL Framework" aims at including the three clusters of response, and associated resource outlay, within a single continuum, resulting in a balanced portfolio of actions across the three clusters by way of an optimization module. It is proposed that this approach be applied separately for each hazard to which the target community is exposed. The author first presented this approach at the Global NAP Expo conference in Songdo, Republic of Korea, on 10 April, 2019.[7]

Figure 1 below illustrates the benefit of the approach for a typical community or region affected by a climate hazard. The unoptimized scenario (top line) shows that gaps in investment in preemptive and contingent arrangements (action gaps) have led to losses that far exceed these gaps. This scenario is, unfortunately, one that characterizes many situations of vulnerability in today's world.  The optimized scenario (bottom line), on the other hand, is the one advocated by the PCL approach; in this scenario, the aggregate value of investment in preemptive and contingent arrangements, as well as of accepted losses, has been minimized.  This has been achieved through increasing investment in both preemptive and contingent action, up to the level that has minimized the total resource outlay.

---

[4] According to the UN Office for the Coordination of Humanitarian Affairs, about half of the humanitarian appeals by the United Nations are being met with corresponding financial contributions. <https://reliefweb.int/sites/reliefweb.int/files/resources/Humanitarian%20Funding%20Update_GHO_30DEC016.pdf>.
[5] Companies have already started to link advances in technology to new approaches to risk management, for example see <www.oneconcern.com>.
[6] Benno Keller, Research Brief: Big Data and Insurance: Implications for Innovation, Competition and Privacy. The Geneva Association – International Association for the Study of Insurance Economics. Undated publication.
[7] <http://napexpo.org/2019/sessions/plenary-1-keynote-presentations-2>

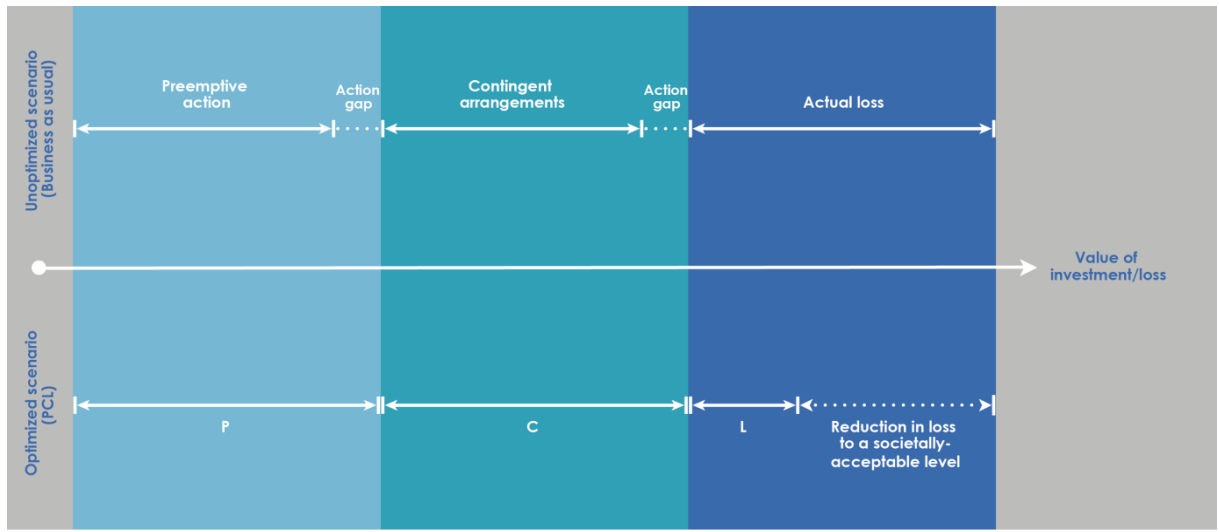

Figure 1:  The PCL approach

One central modality that uniquely characterizes this new approach is that its point of departure – as a prerequisite to the optimization process – is a value-driven consultation with the affected population groups, in which they determine which losses they consider to be tolerable, and which losses they consider to be intolerable.[8]  Each of these two sets of loss categories – the tolerable losses and the intolerable losses – will undergo a different assessment path so that the resulting optimization of actions across all losses takes account of social valuation in addition to economic assessments.  The resulting risk management approach will therefore internalize societal buy-in, since it includes the community's valuation of loss tolerance – effectively giving that community a strong voice in risk management decisions of their elected officials.[9]

## 2.  Description of the methodology

### 2.1 Definition of the clusters

***P: Preemptive action*** – covers the broad range of anticipatory adaptation or risk reduction measures that would be undertaken through planned interventions, including by way of national adaptation plans and any associated subnational or sectoral plans.  These range from soft measures (e.g. regulatory and fiscal incentives, awareness-raising and capacity-building) to concrete adaptation actions (e.g. coastal setbacks, water harvesting and wetland protection);

---

[8] Building on assessment practice in the context of insurance, it is proposed that losses with an annual likelihood higher than 0.5 percent should be considered (e.g. those arising from 1 in 200 year events).
[9] The loss classification process will succeed only with the engagement of representative of all factions of society.  The subsequent social appraisal will further ensure that inequality is not further exacerbated, by applying appropriate coefficients to discount costs and benefits to the rich and magnify them for the poor.

***C: Contingent arrangements*** – are invoked when the impact materializes, and can include mechanisms such as risk transfer, capital market instruments and contingent credit on the one hand, and planned relocation on the other. They can encompass insurance, including parametric insurance, microinsurance and resinsurance, other financial instruments, including bonds and derivatives, reserve funds and other contingent credit arrangements, including the more progressive forecast-based financing,[10] and government guarantees and subsidies;

***L: Loss acceptance*** – in cases where the societally-assessed cost of loss acceptance is less than that of preemptive or contingent action. This largely overlaps with the concept of "risk retention".

**2.2 Assumptions and boundary conditions**

The methodology is intended to optimize resource outlays across the three abovementioned clusters. At its current initial phase of conceptualization, it applies to the perspective of a single administrative level, typically a national or municipal government. This implies that the costs and benefits are assessed from the perspective of that decision-making entity or layer, while fully internalizing societal values as per the consultative assessment process. Future expansions of the methodology can seek to integrate different levels of governance within the same model.

**2.3 A stepwise approach**

While this approach seeks to concurrently evaluate responses across the PCL clusters and optimize implementation as mentioned above, the process also follows a stepwise approach that includes two concentric iterative processes. The first is the internal iterative process – which includes the core optimization process, whereby the PCL optimization module will undertake several iterative assessments in order to finalize the optimization process. The second is the external iterative process, e.g. every five years, which periodically revises the assessment against changes in climate projections, societal priorities, technological advances, actuarial models or economic realities. Figure 2 and the subsequent description lays out how the process would work.

---

[10] <https://www.forecast-based-financing.org/>

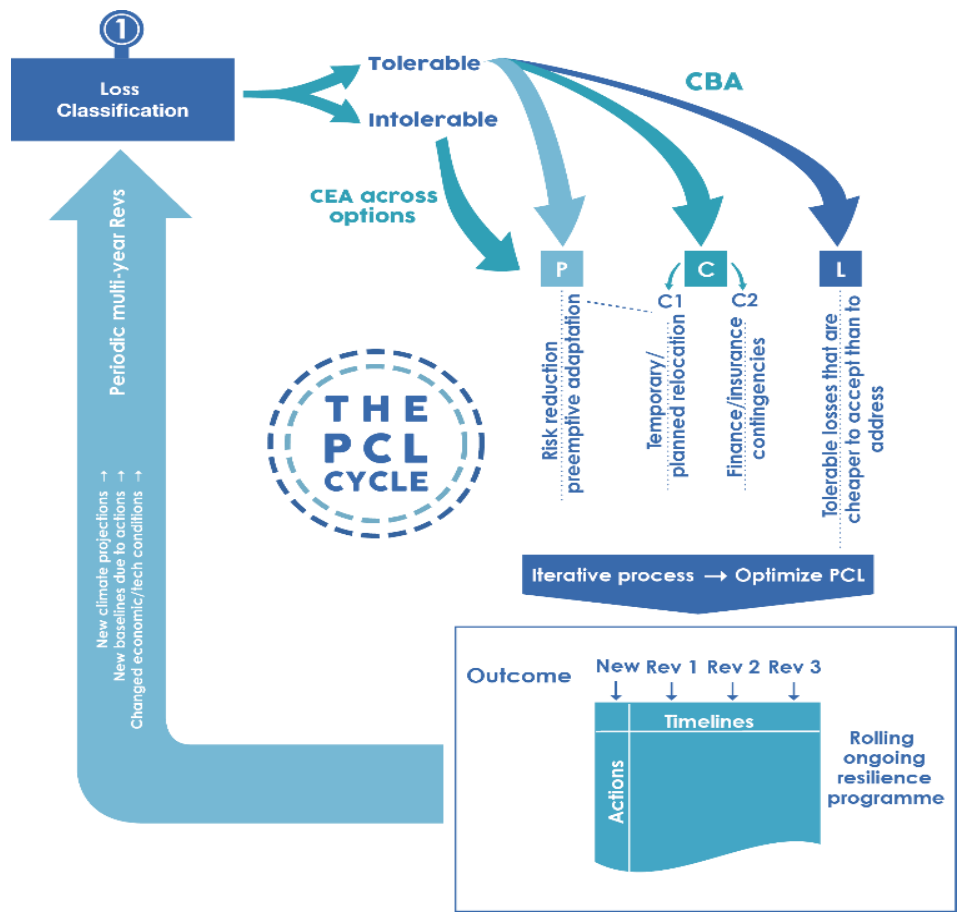

Figure 2: The PCL Cycle

The process of loss classification is based on a societal valuation of losses through a consultative assessment,[11] given that market-based or economic valuation alone would not effectively account for a community's level of tolerability of intangible losses, e.g. those associated with culture, heritage, faith and, most importantly, loss of life.

As shown in the figure above, this step applies a method for gauging societal tolerability of the losses. The input to this step will be a listing of potential losses mapped to risk data, and framed against likely climate change projections. These will include the whole spectrum of losses – human, physical, socioeconomic, sociocultural and environmental. This step will result in listings under two categories of outcomes:

- Intolerable Losses: Those that are not acceptable to society, regardless of the cost of response action: These would include loss of life, loss of national identity, loss of important cultural icons (regardless of their asset value) that contribute to defining the identity of a community, and other losses of particularly high value to that society. The response to these would be addressed exclusively through the "preemptive" action category. Since the losses are intolerable, action must be taken to minimize their risk, even of direct financial costs are higher than benefits. The choice among the different actions that respond to intolerable losses would

---

[11] This would be a representative, inclusive and balanced stakeholder consultation that ensures that all segments of the community are given the opportunity to provide input.

therefore be determined based on cost-effectiveness analysis rather than cost-benefit analysis;[12]
- Tolerable Losses: These would include all other losses and would be classified by way of cost-benefit analysis, into the following three clusters:
    o Preemptive action, where reducing the risk would be deemed to be the least costly option;
    o Contingent action, where ex-post remedial response is deemed to be the least costly action;
    o Loss acceptance, where undertaking preemptive or contingent arrangements is costlier than accepting the losses.

The assessment process to address the two categories is explained in more detail below

*Step 1: Eliminating intolerable losses*

The assessment of risk reduction action to eliminate intolerable losses will also include an identification of any ancillary benefits of such action in reducing the risk of tolerable losses. A determination of the pool of preemptive adaptation actions that would eliminate the intolerable losses will feed into a prioritization and selection of the most desirable preemptive actions, including through a cost-effectiveness analysis or similar approach, rather than a cost-benefit analysis. As mentioned above, the intolerability of the losses under consideration in this step dictates that action be taken to minimize their risk, even if the direct financial benefits outweigh the costs.

This is followed by an assessment of how the selected preemptive risk reduction actions may concurrently reduce any of the other (tolerable) risks as a side benefit, resulting in a revision of the categorized list of potential losses. In other words, this step may remove the need to consider some additional losses (originally deemed to be tolerable) because these will already be addressed by the initial set of preemptive actions that will minimize the intolerable losses.

*Step 2: Addressing tolerable losses*

The outcome of step 1 will provide a revised list of tolerable potential losses that remain after eliminating the intolerable losses. In this step, a process involving multiple iterations is envisioned in order to optimize resource outlays across the P, C and L clusters, using cost benefit analysis (financial, economic and social).[13] The importance of using all three levels of analysis is to ensure that society-wide costs and benefits are considered, and that inequalities and inequities are not exacerbated because of any of the proposed actions.

---

[12] This means that eliminating the risk of such losses will be necessary even if the economic benefit fails to exceed the economic cost. However, an assessment will be needed to choose the most cost-effective course of action towards eliminating the risk.

[13] A financial analysis assesses the inherent costs and benefit of the intervention as a closed system; an economic analysis assesses costs and benefits to the whole economy; and a social analysis uses coefficients to discount costs and benefits to the rich and magnify costs and benefits to the poor, with the objective of avoiding the increase of inequality.

The methodology will also take account of any synergy between actions under the "P" and "C" clusters, in that an intervention under "P" may make a complementary or related intervention under "C" more cost-effective. So, for example, more cost-effective insurance options may be unlocked through additional risk reduction measures, which may in and of themselves not have initially been the preferred option under the "P" cluster. But, coupled with their impact on unlocking actions under the "P" cluster, they become more financially viable.

The approaches described in the Economics of Climate Adaptation,[14] premised on cost-benefit analysis, provide a good approach for prioritizing action to address specific losses in the case of tolerable losses, whether through adaptation/risk reduction, contingency arrangements or loss acceptance.

It should also be noted that existing work, for example the framework proposed by Clarke et al,[15] also provides an effective method for optimizing finance- and insurance-related instruments within the "C" cluster.

The three-step process will be repeated periodically, for example every five years or any other interval preferred by the relevant stakeholders, in order to take account of any change in costs, technologies, impact projections, or societal priorities.

### 3. Differences from existing approaches

In conventional risk management frameworks, cost-benefit analyses or similar assessments typically form a core entry point. While this is an important tool among several that are used under the PCL framework, it is not the main entry point because it excludes societal valuation of the tolerability of losses, and the need to virtually eliminate the risk of intolerable losses, which must be undertaken regardless of whether the benefits exceed the costs. Many legitimate actions which are not necessarily cost-effective would hence be prioritized if they are deemed to be the most effective at minimizing intolerable losses.

Therefore, assessment of tolerability centres around societal valuation rather than on economic or development priorities exclusively. This enhances political, institutional and public buy-in, including for policy reform. Classification of losses based on societal tolerability addresses, in part, the difficulties with dealing with unquantifiable risks, e.g. those relating to ecosystems, mortality, morbidity and culture.

---

[14] http://www.swissre.com/r?19=950&32=10792&7=2529451&40=http%3A%2F%2Fmedia.swissre.com%2Fdocuments%2Frethinking_shaping_climate_resilent_development_en.pdf&41=Download+publication&18=0.2366666006171636

[15] Clarke, Daniel, and Olivier Mahul, Richard Poulter, and Tse-Ling Teh. "Evaluating Sovereign Disaster Risk Finance Strategies: A Framework." World Bank: Disaster Risk Financing and Insurance Program (DRFIP). <http://pubdocs.worldbank.org/en/615151462890267510/DRFI-Clarke-Mahul-Poulter-Teh-Framework-9May16.pdf>

As mentioned above, where cost-benefit or cost-effectiveness analysis is applied within the "P" or "C" clusters under this approach, it will be assessed beyond direct financial appraisal to also include economic and social appraisals.

The methodology envisions dispensing with the notion of "residual risk" in its traditional sense which prioritizes preemptive action regardless of tolerability and irrespective of cost-benefit considerations, and hence it also minimizes any formal dependence on unpredictable levels of humanitarian relief.[16]

The process is iterative, ongoing and long term, in that a broader multi-year iterative process is envisioned in order to re-adjust previous priorities and optimals.  This means that, even if climate projections have not changed over time, the approach can still account for changes in societal perceptions and values, as well as the evolution of technology and comparative pricing between the different response options.

---

[16] Nevertheless, residual "unidentified" risk, which could not have been identified in the assessment phase, will be retained by default.  An example of that would be if a "Category 6" hurricane would ever take place, and which would not have been taken account of in the assessment.